\documentclass[10pt]{article}
\usepackage{latexsym}
\usepackage{eufrak}
\usepackage{euscript}
\usepackage{graphicx}

\title{ A Causal Model of the Hydrogen Atom - New Electron Orbits}

\author{ P.N. Kaloyerou\footnote{Recently retired from the University of Zambia, Physics Dept.}\\ {\small Wolfson College, Linton Road, Oxford OX2 6UD, UK}\\ {\small $<\mathrm{kaloyerou@gmail.com}>$}\\ M. Chiboli and M. Mukutulu\\{\small Department of Physics, School of Natural Sciences,}\\ {\small University of Zambia, PO Box 32379, Lusaka 10101, Zambia} \\ {\small $<\mathrm{mchiboli@gmail.com}>$ and $<\mathrm{mmukutulu@gmail.com}>$} }

\newcommand{\Lv}{\mbox{\boldmath{$L$}}}
\newcommand{\Lvh}{\mbox{\boldmath{$\hat{L}$}}}
\newcommand{\pv}{\mbox{\boldmath{$p$}}}
\newcommand{\vv}{\mbox{\boldmath{$v$}}}
\newcommand{\rv}{\mbox{\boldmath{$r$}}}
\newcommand{\rvs}{\mbox{{\scriptsize\boldmath{$r$}}}}
\newcommand{\er}{\mbox{\boldmath{$\hat{e}_r$}}}
\newcommand{\eth}{\mbox{\boldmath{$\hat{e}_{\theta}$}}}
\newcommand{\eph}{\mbox{\boldmath{$\hat{e}_{\phi}$}}}
\newcommand{\iun}{\mbox{\boldmath{$\hat{i}$}}}
\newcommand{\jun}{\mbox{\boldmath{$\hat{j}$}}}
\newcommand{\kun}{\mbox{\boldmath{$\hat{k}$}}}

\newcommand{\FN}{\mbox{\boldmath{$F_{net}$}}}

\begin{document}

\maketitle
\begin{abstract}
 In this article we develop in detail a  causal model of the hydrogen atom, building on the earlier work of Dewdney and Malik \cite{DM93} in which they outlined a causal model of the hydrogen atom, focusing more on a causal model of angular momentum measurement and of the EPR experiment. We interpret the resulting formulae differently leading to new electron orbits. We develop in detail the relationship between electron orbits, angular momentum and the quantum potential.
\end{abstract}

\section{Introduction\label{INTR}}

Dewdney and Malik in their 1993 paper \cite{DM93} developed a causal model of the measurement of angular momentum and of the EPR experiment \cite{EPR35} based on the de Broglie-Bohm Causal Interpretation \cite{B52, dBR56}. In developing their models,  they first outlined a model of the hydrogen atom.  Here we expand in much more detail the Dewdney and Malik model of the hydrogen atom.  But, we interpret the formula for $L^2$ differently by insisting on two requirements: 
\begin{enumerate}
\item[R1.] Since we are developing a causal model, we require that the angular momentum vector satisfies the classical angular momentum formula  $\Lv=\rv \times \pv$. 
\item[R2.] For full  consistency with the usual Copenhagen interpretation of quantum mechanics, we require that our model be consistent with the quantum mechanical vector model of angular momentum for all hydrogen atom states with magnetic quantum number $m\neq 0.$
\end{enumerate}

Dewdney and Malik only insisted on requirement R1. The $m=0$ state for all values of $l$ is a special case that we will discuss below. As we  shall see,  requirement R1 together with requirement R2  leads to entirely new electron orbits. Specifically,   $m\neq 0$ electron orbits are concentric with  nuclear latitudes, never concentric with the nuclear equator, i.e., electrons never orbit around the nuclear equator. Here and in all that follows, nuclear latitudes and the nuclear equator are defined to lie in planes  orthogonal  to the $z$-axes, whatever its orientation.

Our model therefore differs significantly from the Bohr model of the atom in which electrons orbit the nucleus, i.e., a solar system type model. It also differs from the usual Copenhagen Interpretation (the Born Probability Rule \cite{Born26} and Bohr's Principle of Complementarity (BPC) \cite{BR28}) or its modern offshoots (e.g. information theory), since according to BPC, pictures of electron orbits are viewed as abstractions to aid thought, which cannot be attributed physical reality.  Aside from the quantum postulate, BPC is not a direct interpretation of the mathematical formalism of quantum mechanics\footnote{|See reference  \cite{K16} for a more detail on BPC and the quantum postulate.}, whereas the Born probability rule is an essential interpretational element which links theory with experiment. Therefore, the quantum postulate (which states that,  except for eigenstates, measurement changes the original state of the system) and the Born probability will necessarily be a part of any interpretation of the quantum theory, and certainly a part of the Bohm-de Broglie causal interpretation (CI).

Apart from a different interpretation of the formula for $L^2$, we extend  the work of Dewdney and Malik in the following ways:
\begin{enumerate}
\item[(1)] In addition to the $\psi_{21m}$ hydrogen atom state, we also model the  $\psi_{43m}$ state, both of which are eigenstates of $\hat{\Lvh}$ and $L_z$. 
\item[(2)] We generalise the model to include rotated $\psi_{21m}$ states. 
\item[(3)] We solve the equation of motion for the electron  explicitly for the states $\psi_{nlm}$ and numerically (using  a fourth order Runge-Kutta procedure derived from \cite{BF01}) for  the rotated $\psi_{21m}$ states.
\item[(4)] We consider in detail the relation between electron orbits, the angular momentum vector and the net force (quantum potential plus the electrostatic force).
\end{enumerate}

\section{Brief outline of the Bohm-de Broglie Causal Interpretation}

Substituting $\psi=R(\rv,t)e^{iS(\rvs,t)/\hbar}$ into the Schr\"{o}dinger equation, differentiating and equating real and imaginary terms, gives rise to a continuity equation (CE),
\begin{equation}
\frac{\partial R^2}{\partial t}+\nabla. \left(R^2\frac{\nabla S}{m_p}\right)=0\label{CE},
\end{equation}
and a Hamilton-Jacobi type equation (HJ),
\begin{equation}
-\frac{\partial S}{\partial t}=\frac{(\nabla S)^2}{2m_p}+V+\left(-\frac{\hbar^2}{2m_p}\frac{\nabla^{2}R}{R}\right)\label{HJE}.
\end{equation}
The HJ-equation contains the extra classical term $Q=-\frac{\hbar^2}{2m_p}\frac{\nabla^{2}R}{R}$, which Bohm called the quantum potential. The $R(\rv,t)$ and $S(\rv,t)$-fields are two real fields which co-determine each other.

The continuity equation, Eq. (\ref{CE}), expresses the conservation of probability. As mentioned above, the Born probability rule is an essential element  of any interpretation of the quantum theory. In the causal interpretation, probability enters the interpretation in that initial positions of particles can only be determined with a probability found from the usual probability  density $|\psi|^2=R^2$.

The HJ-equation, Eq. (\ref{HJE}), is interpreted, by analogy with the classical Hamilton-Jacobi equation, as describing particles with kinetic energy
$ KE=\frac{(\nabla S)^{2}}{2m_p}$, potential energy $V(\rv,t)$, total energy 
\begin{equation}
E=-\frac{\partial S}{\partial t},
\end{equation}
momentum
\begin{equation}
\pv=\nabla S, 
\end{equation}
velocity
\begin{equation}
\vv=\frac{\nabla S}{m_p}\label{VEL}
\end{equation}
and equation of motion
\begin{equation}
\vv=\frac{d\rv}{dt}=\frac{\nabla S}{m_p},
\end{equation}
where $m_p$ is the particle mass.

The extra-classical quantum potential, which depends only on the $R$-field, modifies the classical Hamilton-Jacobi equation and gives rise to the quantum behaviour of particles (e.g. interference). Thus, through the HJ-equation,  quantum behaviour can be attributed to the $R$-field. However, since the $R$ and $S$-fields co-determine each other, quantum behaviour of particles can just as well be attributed to the $S$-field.

\section{The equations of the hydrogen atom causal model\label{HACM}}

The hydrogen atom eigenstates of the energy operator, $\hat{H}$,  the angular momentum operator, $\Lvh$, and the $z$-component of the angular momentum operator, $\hat{L}_z$, are given by (see for example \cite{BJ89}, p360)
\begin{equation}
\psi_{nlm}(r,\theta,\phi)=R_{nl}(r)Y_{lm}(\theta,\phi)e^{-iE_{CI}t /\hbar}=N_{sp}R_{nl}(r)P^m_l(\cos\theta)e^{im\phi}e^{-iE_{CI}t /\hbar}, \label{HAST}
\end{equation}
where $R_{nl}$ are  the normalised radial functions, $P^m_l(\cos\theta)$ are the associated Laguerre polynomials, and $N_{sp}$ is a normalisation factor for the spherical Harmonics. The subscripts $n$, $l$ $m$ have their usual meaning as the principle quantum number, the orbital angular momentum quantum number and magnetic quantum number, respectively. We have added the time-dependence, $e^{-iE_nt /\hbar}$,  for  stationary states that is sometimes excluded in text books. 

Setting
\begin{equation}
R(r,\theta)e^{iS(\phi,t)}=N_{sp}R_{nl}(r)P^m_l(\cos\theta)e^{im\phi}e^{-iE_{CI}t /\hbar},
\end{equation}
we get
\begin{equation}
R(r,\theta)=N_{sp}R_{nl}(r)P^m_l(\cos\theta),
\end{equation}
and
\begin{equation}
S[\phi(t),t]=\hbar m\phi(t)-E_{CI} t.\label{SFUN}
\end{equation}
The total energy is given by
\begin{equation}
 -\frac{\partial S}{\partial t}=-\hbar m\frac{d \phi(t)}{d t}+E_{CI}=E_n,\label{STEE}
\end{equation}
where $E_n=\mathrm{constant}$ is given by the Bohr formula for the total energy of an electron in its orbit:
\begin{equation}
E_n=-\frac{\mu}{2\hbar^2}\left( \frac{Z e^2}{4\pi\epsilon_0} \right)^2\frac{1}{n^2},\;\;\;\;\;n=1,2,\dots,\label{ENTOT}
\end{equation}
and where symbols in the formula have their usual meaning: $Z$ is the atomic number, $\mu=\frac{m_eM}{m_e + M}$ is the reduced mass, $e$ is the electron charge and $\epsilon_0$ is the permittivity of free space.  We shall see later, when we solve the equation  for $\phi(t)$,   that it depends only on time, and that 
\[
\frac{d\phi(t)}{dt}=\mathrm{constant}.
\]
Using spherical polar coordinates, the momentum is  given by
\[
\pv=\nabla S=\er\frac{\partial S}{\partial r}+\eth\frac{1}{r}\frac{\partial S}{\partial \theta}+\eph \frac{1}{r \sin\theta}\frac{\partial S}{\partial \phi},
\]
\begin{equation}
\pv=\nabla S=\frac{\hbar m}{r\sin \theta}\eph\label{LMOM}.
\end{equation}
By expressing the unit vector $\eph$ in terms of Cartesian coordinates,
\[
\eph=-\sin\phi\,\iun + \cos\phi\,\jun + 0\,\kun,
\]
then substituting it into Eq. (\ref{LMOM}),  we obtain $\pv$ in terms of Cartesian coordinates:
\begin{equation}
\pv=\frac{\hbar m}{r\sin \theta}(-\sin\phi\,\iun + \cos\phi\,\jun + 0\,\kun).\label{CRLMOM}
\end{equation}
By requirement R1, the angular momentum is given by
\begin{eqnarray}
\Lv&=&\rv\times \pv=\rv\times \nabla S=(\er r)\times\left({\eph}\frac{\hbar m}{r\sin \theta}\right),\nonumber\\
\Lv&=& - \frac{\hbar m}{\sin \theta}\,\eth. \label{AMOM}
\end{eqnarray}
By expressing the unit vector $\eth$ in terms of Cartesian coordinates,
\begin{equation}
\eth=\cos\theta\cos\phi\,\iun + \cos\theta\sin\phi\,\jun - \sin\theta \,\kun,\label{ETHUN}
\end{equation}
then substituting it into Eq. (\ref{AMOM}),  we obtain the following Cartesian components of $\Lv$:
\begin{eqnarray}
L_x&=&-m\hbar  \cot\theta\cos\phi\label{LXC}\\
L_y&=&-m\hbar  \cot\theta\sin\phi\label{LYC}\\
L_z&=& m\hbar \label{LZC}
\end{eqnarray}
From the components of the angular momentum vector and  from Eq. (\ref{AMOM}) we get
\[
L^2=\Lv.\Lv=L_y^2+L_x^2+L_z^2,
\]
\begin{equation}
L^2=\frac{m^2 \hbar^2}{\sin^2\theta}=m^2 \hbar^2\left( \cot^2\theta +1\right). \label{AMSQ}
\end{equation}

 We come now to interpreting formulae (\ref{AMOM}),  (\ref{LZC})  and (\ref{AMSQ}), and it is here that our interpretation begins to differ from that of Dewdney and Malik. The  difference is that we require  formulae  (\ref{AMOM}),  (\ref{LZC}) and (\ref{AMSQ}) to agree with the quantum formalism, i.e., we impose requirement R2. As is well known, the eigenvalues of the operators $\hat{L}_z$ and $\Lvh$  for the hydrogen atom states, Eq. (\ref{HAST}),  are constants of the motion. We therefore require that formulae   (\ref{AMOM}), (\ref{LZC}) and (\ref{AMSQ}) should correctly give the time-independent eigenvalues of these operators.

The formula for $L_z$, Eq. (\ref{LZC}), obviously gives the correct time-independent eigenvalue of $\hat{L}_z$. That the formulae for  $L^2$, (\ref{AMSQ}), gives the correct time-independent eigenvalues of $\hat{L}^2$, at least for the cases $m\neq 0$, follows from the fact that R1 and R2 require that $\theta$ has fixed values given by the formulae:
\begin{eqnarray}
\theta_e&=&\frac{\pi}{2}-\alpha, \;\;\;\;\;m=0,1,2,\ldots\label{THAL1}\\
\theta_e&=&\frac{3\pi}{2}-\alpha, \;\;\;\;\;m=0,-1,-2,\ldots\label{THAL2}\\
\cos\alpha&=&\frac{m}{\sqrt{l(l+1)}}\label{ALAL},
\end{eqnarray}
 which we will derive in \S\ref{EQMPP}. For these values of $\theta$, it is easy to see that formula  (\ref{AMSQ}) gives the  correct eigenvalues of  $\hat{L}^2$:
\[
L^2=l(l+1)\hbar^2.
\]
It therefore follows that formula (\ref{AMOM}) also gives the correct eigenvalues of $\hat{\Lv}$.

Eq. (\ref{LMOM}) shows that  $m=0$ electrons are stationary. It therefore follows, as is also shown by Eqs. (\ref{AMOM}) and (\ref{AMSQ}), that $\Lv$ is zero contrary to the quantum mechanical vector model of angular momentum which attributes the finite values $|\Lv|=\hbar\sqrt{l(l+1)}$ to the angular momentum vector for $m=0$. We note though, that the formula for $L_z$, Eq (\ref{LZC}), gives the correct quantum mechanical  eigenvalues for all values of $m$. 

\section{The electron equations of motion\label{EQMPP}}

Eq. (\ref{VEL}) gives the velocity of a particle. By expressing each side of Eq. (\ref{VEL})  in spherical polar coordinates,
\begin{eqnarray}
\vv&=&\frac{d\rv}{dt}=\frac{d (r\er)}{dt}=\frac{dr}{dt}\er+r\frac{d\theta}{dt}\eth+r\sin\theta\frac{d\phi}{dt}\eph    \label{VSPH}\\
\frac{\nabla S}{m_e}&=&\frac{1}{m_e}\left(\frac{\partial S}{\partial r}\er+\frac{1}{r}\frac{\partial S}{\partial \theta}\eth+\frac{1}{r\sin\theta}\frac{\partial S}{\partial \phi}\eph \right),
\end{eqnarray}
and equating the coefficients of the unit vectors, we obtain the following general equations of motion in spherical polar coordinates:
\begin{equation}
\frac{dr}{dt}=\frac{1}{m_e}\frac{\partial S}{\partial r},\;\;\;\;\;\frac{d\theta}{dt}=\frac{1}{m_e r^2}\frac{\partial S}{\partial \theta},\;\;\;\;\;\frac{d\phi}{dt}=\frac{1}{m_e r^2 \sin^2\theta}\frac{\partial S}{\partial \phi},
\end{equation}
where $m_e$ is the mass of the electron.

Using Eq. (\ref{SFUN}) for $S$, we get the equations of motion for the hydrogen atom electron orbits:
\begin{eqnarray}
\frac{d r}{d t}&=& 0,\;\;\;\;\; r=r_e, \label{DRT}\\
\frac{d \theta}{d t}&= &0,\;\;\;\;\; \theta=\theta_e, \label{DTH} \\
\frac{d \phi}{d t}&=& \frac{m \hbar}{m_e r_e^2 \sin^2\theta_e},\;\;\;\;\; \phi(t)= \frac{m\hbar}{m_e r_e^2 \sin^2\theta_e}\,t+c,  \label{DPHI}
\end{eqnarray}
where $r_e,\theta_e$ and {c} are integration constants.

The constant $r_e$, the distance of the electron from the nucleus, can only be determined with a probability found from the usual radial distribution function $D_{nl}(r_e)=r_e^2 R_{nl}(r_e)^2$. For our computer models, we chose the most probable values of $r_e$ determined from $D_{nl}$, which, as is well known, corresponds to those given by the  Bohr model, which for the  hydrogen atom are
\[
r_e=n^2\frac{m_e}{\mu}a_0,
\]
where $a_0$ is the Bohr radius, $\mu=\frac{Mm_e}{M+m_e}$ is the reduced mass with $M$ the nuclear mass. 

\begin{figure}[h]
\unitlength=1in
\hspace*{1.6in}\includegraphics[width=3.5in,height=3.5in]{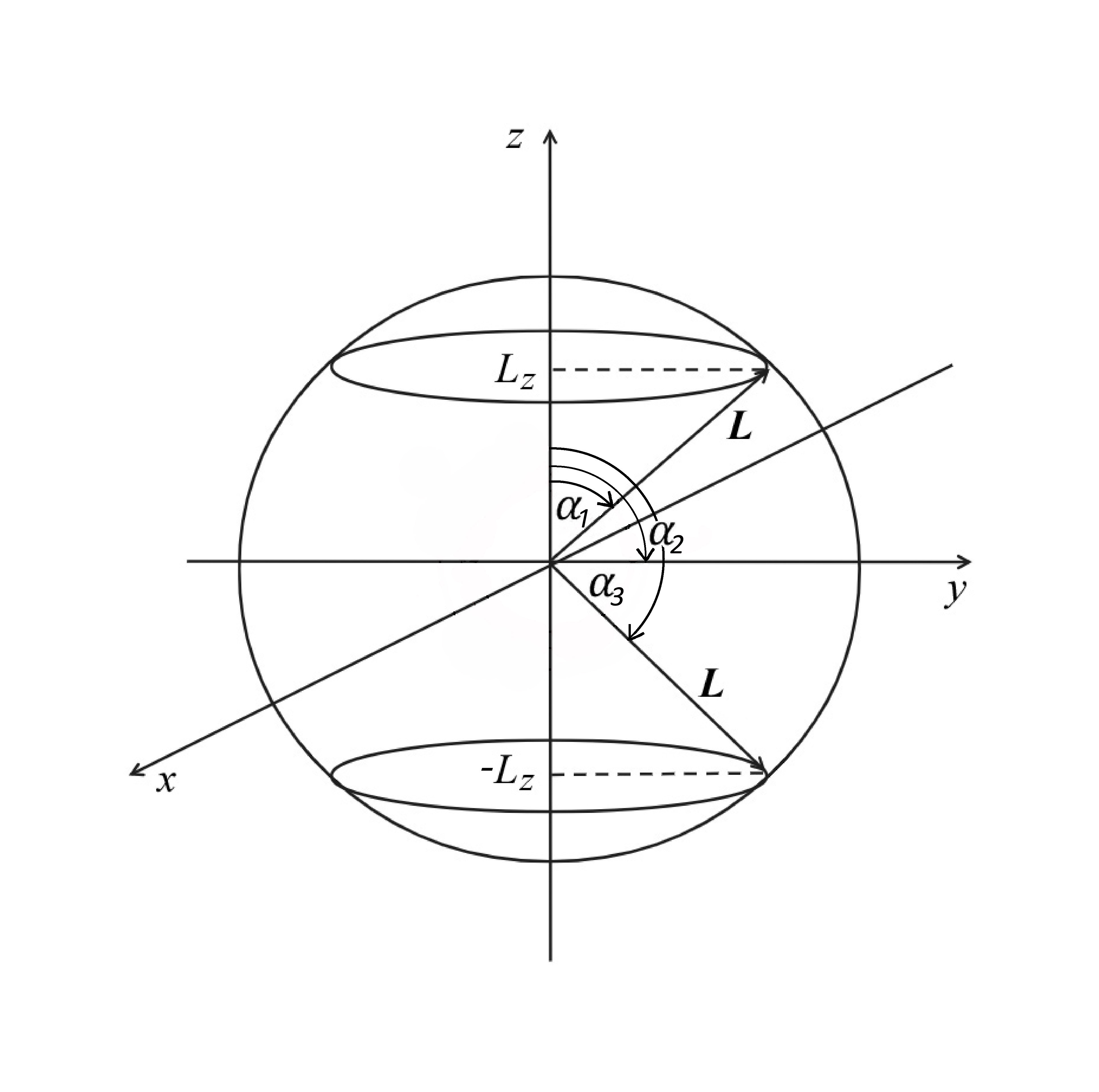}
\caption{Determination of $\alpha$ for $m=-1,0,1$ according to the quantum mechanical vector model of angular momentum.}
\label{HAALPH}
\end{figure}

\begin{figure}[h]
\unitlength=1in
\hspace*{1.6in}\includegraphics[width=3in,height=3.03in]{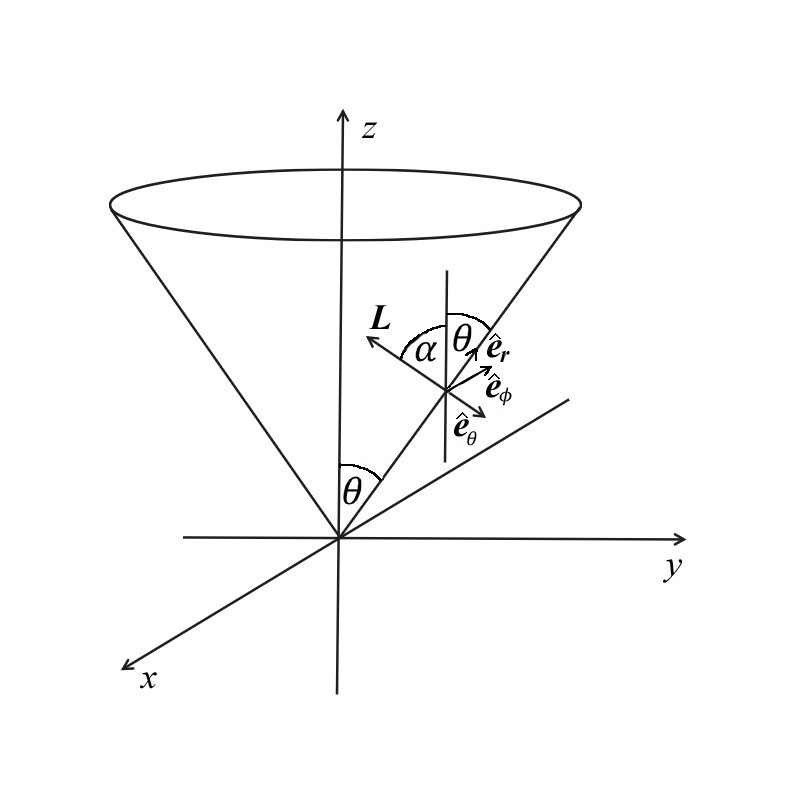}
\caption{Determination of $\theta$ for $l=1$, $m=1$.}
\label{HAAM1}
\end{figure}

\begin{figure}[h]
\unitlength=1in
\hspace*{1.6in}\includegraphics[width=3in,height=3.03in]{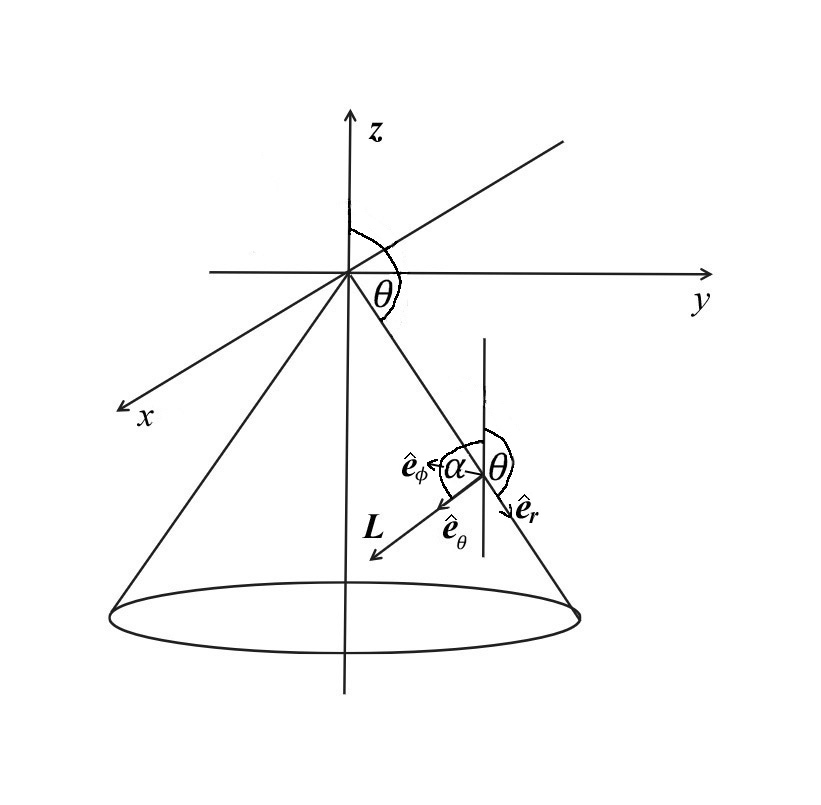}
\caption{Determination of $\theta$ for $l=1$, $m=-1$.}
\label{HAAM2}
\end{figure}

As mentioned, the value of the constant angles $\theta_e$ are determined  by  requirement R2. This is done by first determining the values of $\alpha$ allowed by requirement R2:
\begin{equation}
\cos\alpha=\frac{L_z}{|\Lv|}=\frac{m\hbar}{\sqrt{l(l+1)}\, \hbar}=\frac{m}{\sqrt{l(l+1)}},\;\;\;m=-l,-l+1,-l+2, \ldots, l.\label{PALPH}
\end{equation}
Fig. (\ref{HAALPH})  shows the right angle triangles which determine the allowed  $\alpha$-values for the case $l=1$. Formula (\ref{AMOM}) for $\Lv$ shows that  $\Lv$ is orthogonal to the electron position vector, anticolinear with  $\eth$ for $m=1,2,\ldots, l$ and colinear  with  $\eth$ for $m=-l,-l+1,\ldots,-1$. With these orientations of  $\Lv$ and with $\alpha$ determined by  Eq. (\ref{PALPH}), the angle $\theta_e$ between the $z$-axis and the position vector $\rv$ of the electron  is restricted to the fixed values given by
\begin{equation}
\theta_e=\frac{\pi}{2}-\alpha\;\mathrm{for}\;m=1,2,\ldots, l\;\;\;\;\;\mathrm{and}\;\;\;\;\;\theta_e=\frac{3\pi}{2}-\alpha\;\mathrm{for}\;m=-1,-2,\ldots, l,
\end{equation}
as seen from Figs. (\ref{HAAM1}) and (\ref{HAAM2}).

Substituting Eqs. (\ref{DRT}),  (\ref{DTH}) and  (\ref{DPHI}) into Eq. (\ref{VSPH}) gives the equation of motion of the electron
\begin{equation}
\frac{d\rv}{dt}=\frac{m\hbar}{m_e r_e \sin\theta_e}\,\eph. \label{EEQM}
\end{equation}
To produce the computer models, it is more convenient to express the electron equation of motion in terms of Cartesian coordinates. These can easily be done by expressing both sides of Eq. (\ref{EEQM}) in terms of Cartesian coordinates:
\[
\frac{dx}{dt}\,\iun+\frac{dy}{dt}\,\jun+\frac{dz}{dt}\,\kun=\frac{m\hbar}{m_e r_e \sin\theta_e}\left(-\sin\phi(t)\,\iun+\cos\phi(t)\,\jun+0\,\kun\right),
\]
where we have used $\eph=-\sin\phi(t)\,\iun+\cos\phi(t)\,\jun+0\,\kun$. Equating coefficients of the same unit vector and substituting Eq. (\ref{DPHI}) for $\phi(t)$, we get
\begin{eqnarray}
\frac{dx(t)}{dt}&=&-\frac{m\hbar}{m_e r_e \sin\theta_e}\sin\left(\frac{m\hbar}{m_e r_e^2 \sin^2\theta_e}\,t+c\right),\\
\frac{dy(t)}{dt}&=&\frac{m\hbar}{m_e r_e \sin\theta_e}\cos\left( \frac{m\hbar}{m_e r_e^2 \sin^2\theta_e}\,t+c  \right),\\
\frac{dz(t)}{dt}&=&0,
\end{eqnarray}
These equations are easily integrated to give the time-dependent position vector for the electron from which the electron orbits can be plotted:
\begin{eqnarray}
\rv(t)=x(t)\iun+y(t)\jun+z(t)\kun&=& r_e \sin\theta_e\cos\left(\frac{m\hbar}{m_e r_e^2 \sin^2\theta_e}\,t+c\right)\,\iun\nonumber \\
      &&+ r_e \sin\theta_e\sin\left( \frac{m\hbar}{m_e r_e^2 \sin^2\theta_e}\,t+c  \right) \,\jun \nonumber  \\
      &&+z_0 \, \kun, \label{RVEC}
\end{eqnarray}
where $z_0$ is a constant. We will plot electron orbits for the cases $l=1$, for simplicity, and  $l=3$, as a typical example of higher momentum states.

The equation for $r(t)$ shows that  for the hydrogen atom states given by Eq. (\ref{HAST}),   the electron orbits the nucleus about the $z$-axis in the $xy$-plane. That the electron orbit lies in the $xy$-plane is already indicated by Eq. (\ref{LMOM}) which shows that the linear momentum $\pv$ lies along the unit vector $\eph$ which lies in the $xy$-plane.  

Eq. (\ref{DRT}) shows that $|\rv(t)|=r_e$ is constant so that the vector $\rv(t)$ changes direction with time but not magnitude. Specifically, $\rv(t)$ sweeps out a cone with time as shown in Fig. \ref{HAALPH}. Eq. ( \ref{DTH}) shows  that $\theta$ is also constant. These constant values  determine $z_0$:
\[
z_0=r_e\cos\theta_e
\]
Similarly, the constant values $r_e$ and $\theta_e$, for each value of $m$,  determine the radius $r_0$ of the electron orbit to be $r_0=r_e\sin\theta_e$. Though not obvious to see, the solution of the equations of motion  for $m\neq 0 $ show that the electron orbits are circles of radius $r_0$. The computer models will show this clearly.

That $\theta\neq 90^{\circ}$ for $m\neq 0$, leads to the fundamentally new hydrogen atom electron orbits referred to in the introduction:\\ \\
\hspace*{0.5in}{\bf Electrons orbit the $\mathbf{z}$-axis in orbits concentric to  nuclear latitudes,} \\
\hspace*{0.5in}{\bf never in orbits concentric to the nuclear equator.}\\ \\
We shall see this explicitly in the computer models of \S\ref{ELORBB}. 

We expect this result to extend to all atoms and to arbitrary orientations of the $z$-axis. Indeed, we will show the latter explicitly in the analysis of rotated  hydrogen atom states. This model is therefore contrary to the Bohr model of the atom which envisages electrons orbiting the nucleus much like the planets of our solar system.

Since, as mentioned earlier, $r_e$ can only be determined with a probability  found from the radial distribution function $D_{nl}(r_e)=r_e^2 R_{nl}(r_e)^2$, it follows that $z_0$ can only be determined with the same probability.

The equations of electron motion for the special case $m=0$ become:
\begin{eqnarray}
\frac{dx(t)}{dt}&=&0,\;\;\;\;\;\;x=\mathrm{constant},\\
\frac{dy(t)}{dt}&=&0,\;\;\;\;\;\;y=\mathrm{constant},\\
\frac{dz(t)}{dt}&=&0,\;\;\;\;\;\;z=\mathrm{constant},
\end{eqnarray}
which, as expected from Eq. (\ref{LMOM}) and as mentioned in \S\ref{HACM}, show that $m=0$ electrons are stationary. And, as also pointed out in  \S\ref{HACM}, this contradicts the quantum mechanical vector model of angular momentum, though remains consistent with the classical formula  $\Lv=\rv \times \pv$. As mentioned, the radius of orbits of $m\neq 0$ electrons is given by the radial distribution function,  $D_{nl}(r_e)=r_e^2R_{nl}(r_e)^2$, but since requirement R2 restricts the values of $\theta$ to $\theta_e$, there are positions allowed by the probability density $|\psi_{nlm}|^2$ for which $m\neq 0$ electrons cannot be found. On the other hand,  $m=0$ electrons can be found anywhere around the nucleus consistent with probability density $|\psi_{nl0}|^2$.

\section{Angular momentum}

Substituting Eq. (\ref{DPHI}) for $\phi(t)$ into Eqs. (\ref{LXC}), (\ref{LYC}) and (\ref{LZC}) for the components of the angular momentum vector, gives the equations of motion for the angular momentum vector:
\begin{eqnarray}
L_x(t)&=&-m\hbar  \cot\theta_e\cos\left(   \frac{m\hbar}{m_e r_e^2 \sin^2\theta_e}\,t+c \right)\label{LXCPH}\\
L_y(t)&=&-m\hbar  \cot\theta_e\sin\phi\left(  \frac{m\hbar}{m_e r_e^2 \sin^2\theta_e}\,t+c  \right)\label{LYCPH}\\
L_z&=& m\hbar \label{LZCPH}
\end{eqnarray}
These equations will be used in the computer models to show the relation between the precession of the angular momentum vector and the motion of the electron in its orbit.

\section{Period and frequency of the electron and the angular momentum vector\label{PRFRXX}}

From Eq. (\ref{RVEC}) for $\rv(t)$ and Eqs. (\ref{LXCPH}), (\ref{LYCPH}) and (\ref{LZCPH}) for the components of $\Lv$, we see that the Cartesian components of $\rv$ and $\Lv$ both depend on time in the same way, so that $\Lv$ precesses about the $z$-axis with the same period and frequency as those of the orbiting electron. This period and frequency are given by
\begin{eqnarray}
\phi(t)&=&\frac{m\hbar}{m_e r_e^2\sin^2\theta_e}T+c=2\pi,\nonumber\\
T&=&\frac{(2\pi-c) m_e r_e^2\sin^2\theta_e}{m\hbar},\\
f&=&\frac{1}{T}=\frac{m\hbar}{(2\pi-c) m_e r_e^2\sin^2\theta_e}.
\end{eqnarray}
Eq. (\ref{LMOM}) for $\pv$ shows that the electron orbits in the anticlockwise direction for positive $m$ and clockwise for negative $m$. Since the Cartesian components of $\Lv(t)$ depend on $\phi(t)$ in the same way as the Cartesian components of $\rv(t)$, $\Lv(t)$ will precess about the $z$-axis in the same direction as the electron orbits the $z$-axis; anticlockwise for positive $m$ and clockwise for negative $m$.

Our calculations began  with the classical formula $\Lv=\rv\times\pv$ expressed in terms of spherical polar coordinates.Though obvious, it is useful, by way of checking, to see that the Cartesian form of $\rv$, Eq. (\ref{RVEC}), and of $\pv$, Eq. (\ref{CRLMOM}),  substituted into  $\Lv=\rv\times\pv$ gives the correct Cartesian form of $\Lv$, which a simple calculation, which we will not give,  indeed confirms.

\section{The quantum potential and the net force\label{QPNETF}}

 The quantum potential $Q$ is best found by rearranging the HJ-equation, Eq. (\ref{HJE}):
\begin{equation}
Q= -\frac{\partial S}{\partial t}-\frac{(\nabla S)^2}{2m_e}-V,\label{QPF}
\end{equation}
where $V$ is the spherical symmetric hydrogen atom electrostatic potential given by the Bohr formula:
\begin{equation}
V=-\frac{\mu}{\hbar^2}     \left(\frac{Ze^2}{4\pi\epsilon_0}\right)^2       \frac{1}{n^2}= - \frac{Ze^2}{(4\pi\epsilon_0)r}.\label{HPOT}
\end{equation}
The net force is given by the sum of the negative gradients of the classical and quantum potentials:
\begin{eqnarray}
\FN&=&-\nabla Q + (-\nabla V) =-\nabla\left[-\frac{\partial S}{\partial t}-\frac{(\nabla S)^2}{2m_e}\right]+(\nabla V) + (-\nabla V), \nonumber\\
\FN&=&-\nabla\left[-\frac{\partial S}{\partial t}-\frac{(\nabla S)^2}{2m_e}\right]\label{FFNETT}.
\end{eqnarray}
We see that the quantum force exactly balances the classical electrostatic force. Using Eq. (\ref{STEE}) for $-\frac{\partial S}{\partial t}$, we get
\[
 \nabla\left(-\frac{\partial S}{\partial t}\right)=\nabla E_n=0,\
\]
since $E_n=\mathrm{constant}$.

Substituting this result into Eq. (\ref{FFNETT}), we get
\begin{equation}
\FN=\nabla\left[ \frac{(\nabla S)^2}{2m_e}\right]. \label{FNTTZ}
\end{equation}
Using Eq. (\ref{LMOM}) we have
\[
\FN=\nabla\left[ \frac{(\nabla S)^2}{2m_e}\right]=\frac{\hbar^2 m^2}{2 m_e}  \left[\frac{1}{\sin^2\theta}\frac{\partial \,r^{-2}}{\partial r}\er+\frac{1}{r^3}\frac{\partial \sin^{-2}\theta}{\partial \theta}\eth+0\,\eph \right],
\]
so that  the net force becomes
\begin{equation}
\FN=-\frac{m^2\hbar^2}{m_e r^3\sin^2\theta}(\er+\cot\theta\,\eth).\label{FFNET}
\end{equation}
Substituting Eq. (\ref{ETHUN}) for $\eth$ and
\[
\er=\sin\theta \cos\phi \,\iun +\sin\theta\sin\phi\,\jun+\cos\theta\,\kun
\]
into Eq. (\ref{FFNET}), we get $\FN$ in terms of Cartesian coordinates:
\begin{equation}
\FN=-\frac{m^2\hbar^2}{m_e r^3\sin^3\theta}\,(\cos\phi\,\iun+\sin\phi\,\jun).\label{FNET}
\end{equation}
We see that $\FN$ lies in the same $xy$-plane as the electron orbit, pointing toward the $z$-axis as required  for the centripetal force needed to produce the electron circular orbits. The net force acting on electrons in their fixed orbits is therefore given by substituting $\theta=\theta_c$, $r=r_e$ and $\phi=\phi(t)$ into Eq. (\ref{FNET}):
\begin{equation}
\FN=-\frac{m^2\hbar^2}{m_e r_e^3\sin^3\theta_e}\,[\cos\phi(t)\,\iun+\sin\phi(t)\,\jun].\label{FNETEE}
\end{equation}

For an electron of linear momentum $\pv$ given by Eq. (\ref{LMOM}) to move in circular orbit of radius $r_0=r_e\sin\theta_e$, the magnitude of the  required centripetal force is
\begin{equation}
|F_c|=\frac{m_e v^2}{r_e\sin\theta_e}=\frac{m_e}{r_e\sin\theta_e}\left(\frac{ \pv.\pv}{m_e^2}\right)=\frac{m^2\hbar^2}{m_e r_e^3\sin^3\theta_e}=|\FN|.\label{FCEP}
\end{equation}
We see, therefore, that the net force is just the centripetal force needed for the electron circular orbits.

We note, however, that the net force is zero for $m=0$ so that the quantum force exactly balances the force due to the electrostatic potential. This result is, of course, necessary for consistency since the $m=0$ electron is stationary.

\section{Energy}

For the total electron energy we neglect fine structure and relativistic corrections, and use the total energy as calculated by the Bohr formula (\ref{ENTOT}). . For  $n=2$, $l=1$, we have
\[
-\frac{\partial S}{\partial t}=-5.447\times10^{-19}\;\mathrm{J}.
\]
As shown by  Eq.  (\ref{STEE}), for $m\neq0$ $\phi(t)$ contributes to the total energy, in fact, the $\phi(t)$ contribution is almost equal to the total energy, so  that the value of $E_{CI}$ is very small. For $n=2$, $l=1$, we have 
\begin{eqnarray}
-m\hbar\frac{d\phi(t)}{dt}&=&-m\hbar\left(\frac{m\hbar}{m_e r_e^2\sin^2{45^{\circ}}}\right)=-5.444\times 10^{-19}\;\mathrm{J},\;\;\;\mathrm{for}\;m=\pm 1,\nonumber\\
E_{CI}&=&-0.003\times 10^{-19},\;\;\;\mathrm{for}\;m=\pm 1.\nonumber
\end{eqnarray}
Because of the smallness of $E_{CI}$ values for  $m\neq 0$, in this section and in this section only, we  give numerical values to three decimal places . Eq.  (\ref{STEE}) shows that for $m=0$ the $\phi(t)$ term does not contribute to the total energy, so that
\[
-\frac{\partial S}{\partial t}=E_{CI}=E_n=-5.447\times10^{-19}\;\mathrm{J}, \;\;\;m=0.
\]

For $m\neq0$ electrons, the  kinetic energy  given by $KE_{CI}=\frac{(\nabla S)^2}{2m_e}$ is of the same order of magnitude  as that given by the Bohr formula
\[
KE_{Bohr}= \frac{\mu}{2\hbar^2}\left( \frac{Z e^2}{4\pi\epsilon_0} \right)^2\frac{1}{n^2},\;\;\;\;\;n=1,2,\dots,
\]
but not necessarily identical.  For example, for the case $Z=1$, $n=2$, $l=1$, $KE_{CI}=2.722\times 10^{-19}$ J, while $KE_{Bohr}=5.447\times 10^{-19}$ J, so that $KE_{CI}$ is essentially half as big as $KE_{Bohr}$. As mentioned, for $m=0$ electrons $KE_{CI}=0$.  Since we are concerned with the hydrogen atom, we take $Z=1$ in all numerical examples that follow.

The potential energy that appears  in the HJ-equation for $Z=1$ is found using Eq. (\ref{HPOT}): $V=-10.894\times 10^{-19}$ J.

The magnitude of the quantum potential can be found from Eq. (\ref{QPF}). For $n=2$, $l=1$ and $m=\pm 1$, we have
\[
Q=2.725\times 10^{-19}\;\textrm{J}, 
\]
while  for $n=2$, $l=1$ and $m=0$ we  have
\[
Q=5.447\times 10^{-19}\;\textrm{J}.
\]
\section{Electron orbits about an arbitrary rotated axis}

Thus far, we have considered eigenstate of $\Lvh$ and $\hat{L}^2$ giving rise to electron orbits around the $z$-axis. We now want to consider more general states which give rise to electron orbits about an arbitrary axis. To do this, we will consider the specific states  $\psi_{21m}$ as representative examples, and, for simplicity, consider rotations of the coordinate axes by an angle $\beta$ clockwise  about the $y$-axis (with axes orientated as in Fig. \ref{HAALPH}). 

The states $\psi_{21m}$ are given by
\begin{eqnarray}
\psi_{211} &=&-AR_{21}(r)\sin(\theta)e^{i\phi}=-AR_{21}\sin\theta(\cos\phi+i\sin\phi),   \label{OHAST1}\\
\psi_{210} &=&AR_{21}(r)\sqrt{2}\cos\theta,   \label{OHAST2}\\
\psi_{21,-1} &=&AR_{21}(r)\sin(\theta)e^{-i\phi}=AR_{21}\sin\theta(\cos\phi-\sin\phi),   \label{OHAST3}
\end{eqnarray}
where $A=\left(\frac{3}{8\pi}\right)^{1/2}$, 
$
R_{21}(r)=\frac{1}{3^{1/2}}\left(\frac{1}{2a_{\mu}}\right)^{3/2}\left(\frac{r}{a_{\mu}}\right) e^{-r/(2a_{\mu})},
$
and where $a_{\mu}=\frac{\mu(4\pi\epsilon_0)\hbar^2}{m_e^2 e^2}$ is the modified Bohr radius. To obtain the wave functions in terms of the rotated coordinates, we first want to express the above wave functions in terms of $x$, $y$ and $z$.  We do this by rearranging the expressions for Cartesian coordinates in terms of spherical polar coordinates,
\[
\sin\theta\cos\phi=\frac{x}{r},\;\;\;\;\;\sin\theta\sin\phi=\frac{y}{r},\;\;\;\;\; \cos\theta=\frac{z}{r},
\]
then substituting these into the wave functions (\ref{OHAST1}), (\ref{OHAST2}) and (\ref{OHAST3}). We get
\begin{eqnarray}
\psi_{211} &=&-\frac{AR_{21}(r)}{r}(x+iy),    \label{ZHAST1}\\
\psi_{210} &=&\frac{AR_{21}(r)\sqrt{2}}{r}z,     \label{ZHAST2}\\
\psi_{21,-1} &=&\frac{AR_{21}(r)}{r}(x-iy).  \label{ZHAST3}
\end{eqnarray}
Substituting the expressions for $x$, $y$ and $z$ in terms of the rotated coordinates  $x'$. $y'$ and $z'$, i.e., 
\[
x=z'\sin\beta+x'\cos\beta,\;\;\;\;\;z=z'\cos\beta-x'\sin\beta,\;\;\;\;\; y=y',
\]
into Eqs. (\ref{ZHAST1}), (\ref{ZHAST2}) and (\ref{ZHAST3}) we obtain the expression of the  $\psi_{21m}$ states in terms of the rotated coordinates:
\begin{eqnarray}
\psi_{211} &=&-\frac{AR'_{21}(r')}{r'}( z'\sin\beta+x'\cos\beta   +iy'),    \label{RAHAST1}\\
\psi_{210} &=&\frac{AR'_{21}(r')\sqrt{2}}{r'}( z'\cos\beta-x'\sin\beta  ),     \label{RAHAST2}\\
\psi_{21,-1} &=&\frac{AR'_{21}(r')}{r'}(z'\sin\beta+x'\cos\beta -iy'   ). \label{RAHAST3}
\end{eqnarray}
These states are linear combinations of  the eigenstates $\psi'_{21m}$ of the operators $\hat{L}_{z'}$ and $\Lvh'$ of the rotated coordinate system:
\begin{eqnarray}
\psi_{211} &=&\psi'_{211}\left(\frac{1+\cos\beta}{2}\right)-\psi'_{210}\frac{\sin\beta}{\sqrt{2}}+\psi'_{21,-1}\left(\frac{1-\cos\beta}{2}\right),   \label{rotwf1}\\
\psi_{210} &=&\psi'_{211}\frac{\sin\beta}{\sqrt{2}}+\psi'_{210}\cos\beta+\psi'_{21,-1}\frac{\sin\beta}{\sqrt{2}},    \label{rotwf2}\\
\psi_{21,-1} &=&\psi'_{211}\left(\frac{1-\cos\beta}{2}\right)+\psi'_{210}\frac{\sin\beta}{\sqrt{2}}+\psi'_{21,-1}\left(\frac{1+\cos\beta}{2}\right).  \label{rotwf3}
\end{eqnarray}
 
\subsection{The equations of motion in the rotated frame\label{EQMRFZZ}}
The electron equation of motion in the rotated frame is
\begin{equation} 
\frac{d \rv'(t)}{dt}=\frac{\nabla S_{21m}(x',y',z')}{m_e},
\end{equation}
or, in component form, 
\begin{equation} 
\frac{d x'(t)}{dt}=\frac{1}{m_e}\frac{\partial S_{21m}}{\partial x'},\;\;\;\;\;\frac{d y'(t)}{dt}=\frac{1}{m_e}\frac{\partial S_{21m}}{\partial  y'},\;\;\;\;\;\frac{d z'(t)}{dt}=\frac{1}{m_e}\frac{\partial S_{21m}}{\partial z'}.\label{REQM}
\end{equation}
The formula for $S$ is given by
\begin{equation} 
S_{21m}=\frac{\hbar}{2i}\ln\left(\frac{\psi_{21m}}{{\psi^*}_{21m}} \right)=\frac{\hbar}{2i}\ln\left(\frac{\phi_{21m}}{{\phi^*}_{21m}} \right),\label{FORSS}
\end{equation}
where 
\begin{equation}
\phi_{21m}=\psi_{21m}\left(\frac{r'}{AR'_{21}(r')}\right),\label{PHIXX}
\end{equation}
with $\psi_{21m}$ given by Eqs. (\ref{RAHAST1}), (\ref{RAHAST2}) and (\ref{RAHAST3}). The derivative of $S_{21m}$ with respect to $x'$ is given by
\begin{equation} 
\frac{  \partial S_{21m} }{\partial x'}=\frac{\hbar}{2i} \left(   \frac{1}{\phi_{21m}} \frac{\partial \phi_{21m}}{\partial x'}-  \frac{1}{{\phi^*}_{21m}}  \frac{   \partial {\phi^*}_{21m}   }{\partial x'}  \right),\label{DSDXBB}
\end{equation}
with similar formula for $y'$ and $z'$.

The partial derivatives of $S_{21m}$ with respect to the rotated coordinates, $x',y',z'$,  lead to the following equations of motion:
\begin{eqnarray}
\frac{dx'_1}{dt}&=&\frac{1}{m_e}\frac{\partial S_{211}}{\partial x'_1}=  -\frac{\hbar}{m_e}. \frac{ y'_1(t)\cos\beta}{ \phi_{211}.{\phi^*}_{211} },\;\;\;\;
\frac{dy'_1}{dt}=\frac{1}{m_e}\frac{\partial S_{211}}{\partial y'_1}= \frac{\hbar}{m_e}\left[ \frac{z'_1(t)\sin\beta+x'_1(t)\cos\beta}{ \phi_{211}.{\phi^*}_{211} }\right],\nonumber\\
\frac{dz'_1(t)}{dt}&=&\frac{1}{m_e}\frac{\partial S_{211}}{\partial z'_1}=  -\frac{\hbar}{m_e}\frac{y'_1(t)\sin\beta}{ \phi_{211}.{\phi^*}_{211} },\nonumber\\
\frac{dx'_0(t)}{dt}&=&\frac{1}{m_e}\frac{\partial S_{210}}{\partial x'_0}=0,\;\;\;\frac{dy'_0(t)}{dt}=\frac{1}{m_e}\frac{\partial S_{210}}{\partial y'_0}=0,\;\;\;\frac{dz'_0(t)}{dt}=\frac{1}{m_e}\frac{\partial S_{210}}{\partial z'_0}=0,\label{RTM0}\\
\frac{dx'_{-1}(t)}{dt}&=&\frac{1}{m_e}\frac{\partial S_{21,-1}}{\partial x'_{-1}}=   \frac{\hbar}{m_e}.\frac{y'_{-1}(t)\cos\beta}{ \phi_{21,-1}.{\phi^*}_{21,-1} },\;\;\;\;
\frac{dy'_{-1}}{dt}=\frac{1}{m_e}\frac{\partial S_{21,-1}}{\partial y'_{-1}}=  - \frac{\hbar}{m_e}\left[\frac{ z'_{-1}(t)\sin\beta+x'_{-1}(t)\cos\beta}{ \phi_{21,-1}.{\phi^*}_{21,-1} }\right],\nonumber\\
\frac{dz'_{-1}}{dt}&=&\frac{1}{m_e}\frac{\partial S_{21,-1}}{\partial z'_{-1}}=   \frac{\hbar}{m_e}\frac{y'_{-1}(t)\sin\beta}{ \phi_{21,-1}.{\phi^*}_{21,-1} },\nonumber
\end{eqnarray}
 Note that $\phi_{211}.{\phi^*}_{211}=\phi_{21,-1}.{\phi^*}_{21,-1}$. It is these resulting equations of motion that we will use for the computer models of the rotated electron orbits.  Eq. (\ref{RTM0}) confirms the obvious result that the $m=0$ electron remains stationary.

As mentioned in \S\ref{INTR}, the equations of motion  in the rotated frame are solved numerically using a fourth order Runge-Kutta procedure. Obviously, for consistency with the quantum mechanical vector model, the orbits must still be fixed by angle $\theta_e$ relative to the rotated $z$-axis. This is achieved  by choosing the initial position of the electron in each orbit in the Runge-Kutta procedure to correspond to the  initial position in the unrotated orbit by using the following inverse transformations:
\begin{eqnarray}
x'=x\cos \beta -z\sin\beta,\nonumber\\
y'=y,\nonumber\\
z'=x\sin\beta+ z\cos\beta\nonumber.
\end{eqnarray}
We will list the initial values used in the Runge-Kutta procedure in Table 2 of \S\ref{ELOR}.

\section{The quantum potential and net force in the rotated frame\label{QPNETFROT}}

The net force in the rotated frame is given by Eq. (\ref{FNTTZ}) expressed in terms of the Cartesian coordinates of the rotated frame:
\begin{eqnarray}
\FN_{(21m)}=\nabla\left[ \frac{(\nabla' S_{21m})^2}{2m_e}\right]&=&\frac{1}{m_e}\left[\iun \left(\frac{\partial^2 S_{21m}}{\partial x'^2}\frac{\partial S_{21m}}{\partial x'}+  
\frac{\partial^2 S_{21m}}{\partial x' \partial y'}\frac{\partial S_{21m}}{\partial y'} + \frac{\partial^2 S_{21m}}{\partial x' \partial z'}\frac{\partial S_{21m}}{\partial z'}  \right)\right.  \nonumber\\
&+&\jun\left(\frac{\partial^2 S_{21m}}{\partial y'\partial x'}\frac{\partial S_{21m}}{\partial x'}+  
\frac{\partial^2 S_{21m}}{\partial y'^2}\frac{\partial S}{\partial y'} + \frac{\partial^2 S_{21m}}{\partial y' \partial z'}\frac{\partial S_{21m}}{\partial z'}  \right)  \nonumber\\
&+&\kun  \left.   \left(\frac{\partial^2 S_{21m}}{\partial x'\partial z'}\frac{\partial S_{21m}}{\partial x'}+  
\frac{\partial^2 S_{21m}}{\partial y' \partial z'}\frac{\partial S_{21m}}{\partial y'} + \frac{\partial^2 S_{21m}}{\partial z'^2}\frac{\partial S}{\partial z'}  \right)\right]. \label{FNETZZ}  
\end{eqnarray}
Formula (\ref{DSDXBB}) gives the first derivative of $S_{21m}$ with respect to $x'$, while the second derivative of $S$ with respect to $x'$ is given by
\begin{equation}
\frac{\partial^2 S_{21m}}{\partial x'^2}=\frac{\hbar}{2i}\left[ -\frac{1}{\phi^{2}_{21m}}\left(\frac{\partial \phi_{21m}}{\partial x'}\right)^2 +\frac{1}{\phi_{21m}}\frac{\partial^2 \phi_{21m}}{\partial x'^2} +  \frac{1}{\phi^{*2}_{21m}}\left(\frac{\partial \phi^{*}_{21m}}{\partial x'}\right)^2-\frac{1}{\phi^{*}_{21m}}\frac{\partial^2 \phi^{*}_{21m}}{\partial x'^2}\right],\nonumber
\end{equation}
with similar formulae for derivatives with respect to $y'$ and $z'$.  The formulae for mixed derivatives is exemplified by  the derivative of  $S_{21m}$ with respect to $x'$ and $y'$:
\[
 \frac{\partial^2 S_{21m}}{\partial x' \partial y'}=\frac{\hbar}{2i}\left[ -\frac{1}{\phi^{2}_{21m}}\frac{\partial \phi_{21m}}{\partial x'}\frac{\partial \phi_{21m}}{\partial y'}       +\frac{1}{\phi_{21m}}\frac{\partial^2 \phi_{21m}}{\partial x'\partial y'} +  \frac{1}{\phi^{*2}_{21m}}\frac{\partial \phi^{*}_{21m}}{\partial x'}\frac{\partial \phi^{*}_{21m}}{\partial y'}-\frac{1}{\phi^{*}_{21m}}\frac{\partial^2 \phi^{*}_{21m}}{\partial x'\partial y'}\right].
\] 
The derivatives of  $S_{21m}$ with respect to $x',z'$ and $y',z'$  are similar to the above.

Substituting Eq. (\ref{PHIXX}) for $\phi_{210}$,  with Eq.(\ref{RAHAST2}) for $\psi_{210}$, in formula (\ref{FORSS}) gives
\[
S_{210}=\frac{\hbar}{2i}\ln\left(\frac{\phi_{210}}{\phi^{*}_{210}} \right)=\frac{\hbar}{2i}\ln\left(\frac{\phi_{210}}{\phi_{210}} \right)= \frac{\hbar}{2i}\ln\left(1 \right)=0,
\]
from which it follows that all derivatives of $S_{210}$ are zero. Hence, $\FN_{(210)}=0$ as expected, since the $m=0$ electron is stationary.

Applying the above formulae to the states $\phi_{21,\pm 1}$ leads to lengthy and tedious differentiations and lengthy final expressions for $\FN_{(21,\pm 1)}$. Therefore, instead of giving the final expressions for  $\FN_{(21,\pm 1)}$, we list the  derivatives of $S_{21,\pm 1}$, which, when  substituted into Eq. (\ref{FNETZZ}), give $\FN_{(21,\pm 1)}$:

\begin{eqnarray}
 \frac{\partial S_{211}}{\partial x'}&=& -\frac{\hbar y' \cos\beta}{\phi_{211}.\phi^{*}_{211}},\nonumber\\
 \frac{\partial S_{211}}{\partial y'}&=& \frac{\hbar \left( z' \sin\beta+x\cos\beta\right)}{\phi_{211}.\phi^{*}_{211}},\nonumber\\
 \frac{\partial S_{211}}{\partial z'}&=& -\frac{\hbar y' \sin\beta}{\phi_{211}.\phi^{*}_{211}},\nonumber\\
 \frac{\partial^2 S_{211}}{\partial x'^2}&=& \frac{2\hbar \left(    y'z' \cos^2\beta\sin\beta+x'y'\cos^3\beta\right)  }{\left(\phi_{211}.\phi^{*}_{211}\right)^2},\nonumber\\
 \frac{\partial^2 S_{211}}{\partial y'^2}&=& \frac{2\hbar \left(    y'z' \sin\beta+x'y'\cos\beta\right)  }{\left(\phi_{211}.\phi^{*}_{211}\right)^2},\nonumber\\
 \frac{\partial^2 S_{211}}{\partial z'^2}&=& \frac{2\hbar \left(    y'z' \sin^3\beta+x'y'\cos\beta\sin^2\beta\right)  }{\left(\phi_{211}.\phi^{*}_{211}\right)^2},\nonumber\\
\frac{\partial^2 S_{211}}{\partial x'\partial y'}&=& - \frac{ \hbar  \left(z'^2\cos\beta \sin^2\beta+x'^2\cos^3\beta - y'^2 \cos\beta + 2x'z'\cos^2\beta\sin\beta \right) }{\left(\phi_{211}.\phi^{*}_{211}\right)^2},\nonumber\\
\frac{\partial^2 S_{211}}{\partial x'\partial z'}&=&   \frac{ 2\hbar \left(y'z'\cos\beta \sin^2\beta+x'y'\cos^2\beta\sin\beta \right) }{\left(\phi_{211}.\phi^{*}_{211}\right)^2},\nonumber\\
\frac{\partial^2 S_{211}}{\partial y'\partial z'}&=& - \frac{ \hbar  \left(z'^2\sin^3\beta+x'^2\cos^2\beta\sin\beta - y'^2 \sin\beta + 2x'z'\cos\beta\sin^2\beta \right) }{\left(\phi_{211}.\phi^{*}_{211}\right)^2},\nonumber
\end{eqnarray}
where
\begin{eqnarray}
\phi_{211}.\phi^{*}_{211}&=&z'^2\sin^2\beta+x'^2\cos^2\beta+y'^2+2x'z'\sin\beta\cos\beta,\nonumber\\
\left(\phi_{211}.\phi^{*}_{211}\right)^2&=&z'^4\sin^4\beta+2x'^2z'^2\cos^2\beta\sin^2\beta+2y'^2z'^2\sin^2\beta +4x'z'^3\sin^3\beta\cos\beta+x'^4\cos^4\beta \nonumber\\
&+&2x'^2y'^2 \cos^2\beta+ 4x'^3z\sin\beta\cos^3\beta+ y'^4 + 4x'y'^2z'\sin\beta\cos\beta + 4x'^2z'^2   \sin^2\beta\cos^2\beta.\nonumber
\end{eqnarray}

The derivatives of  $S_{21,-1}$ are just the negatives of those for $S_{211}$, while $\phi_{21,-1}.\phi^{*}_{21.-1}=\phi_{211}.\phi^{*}_{211}$, as mentioned earlier.

\section{Angular momentum of the rotated electron orbits}

The angular momentum in Cartesian coordinates in the rotated frame for the states $ \psi_{21m}$ is given by
\begin{equation}
\Lv_{21m}=\left(y' \frac{\partial S_{21m}}{\partial z'}-z\frac{\partial S_{21m}}{\partial y'}  \right)\iun -  \left(x' \frac{\partial S_{21m}}{\partial z'}-z'\frac{\partial S_{21m}}{\partial x'}  \right)\jun + \left(x' \frac{\partial S_{21m}}{\partial y'}-y'\frac{\partial S_{21m}}{\partial x'}  \right)\kun.\label{ANGMXZ}
\end{equation}
Substituting the derivatives of $S_{21m}$ into Eq. (\ref{ANGMXZ}), we get
\begin{eqnarray}
\Lv_{211}(t)\;&=&\!\!\!\frac{\hbar}{\phi_{211}.\phi^{'*}_{211}}\left\{-[\{y'(t)^2+z'(t)^2\}\sin\beta  +x'z'\cos\beta]\,\iun+  [ x'(t)y'(t)\sin\beta-y'(t)z'(t)\cos\beta]\,\jun \right. \nonumber\\
&&\left.+ [ x'(t)z'(t)\sin\beta+\{x'(t)^2+y'(t)^2\}\cos\beta       ]\,\kun \right\}\nonumber\\
\Lv_{21,-1}(t)\!\!\!&=&-\Lv_{211}(t).\label{AM11}\\
\L'_{210}(t)\;&=&0. \nonumber
\end{eqnarray}

\begin{figure}[h]
\unitlength=1in
\hspace*{1.6in}\includegraphics[width=4in,height=3.7in]{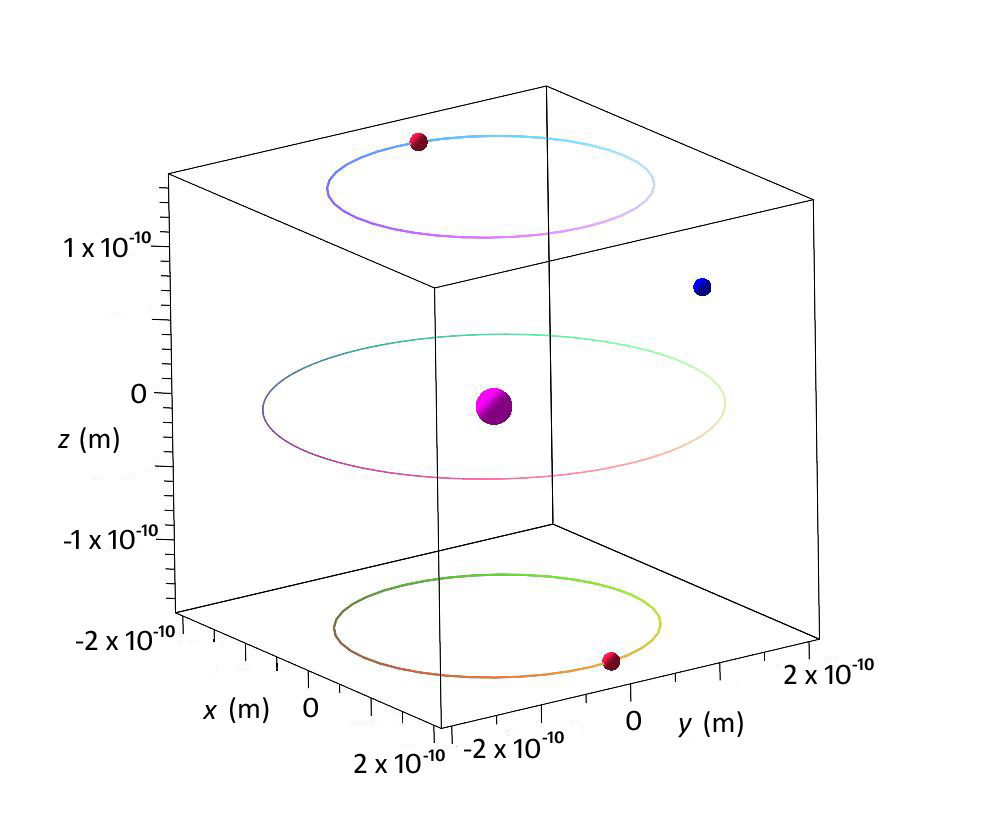}
\caption{Possible hydrogen atom electron orbits for the case $n=2$, $l=1$.}
\label{ORB3}
\end{figure}

\begin{figure}[h]
\unitlength=1in
\hspace*{1.6in}\includegraphics[width=4in,height=4.3in]{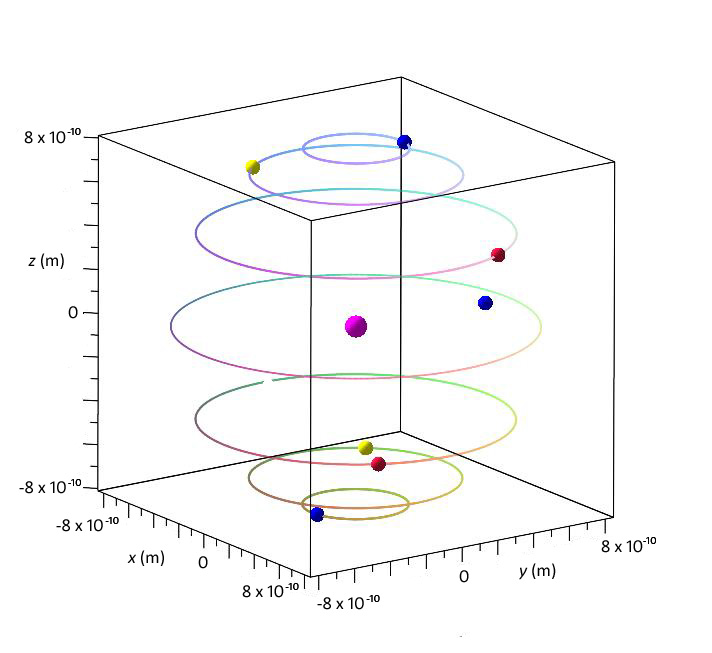}
\caption{Possible hydrogen atom electron orbits for the case $n=4$, $l=3$.}
\label{ORB7}
\end{figure}
\vspace*{0.4cm}

\begin{figure}[h]
\unitlength=1in
\hspace*{1.6in}\includegraphics[width=4in,height=3.8in]{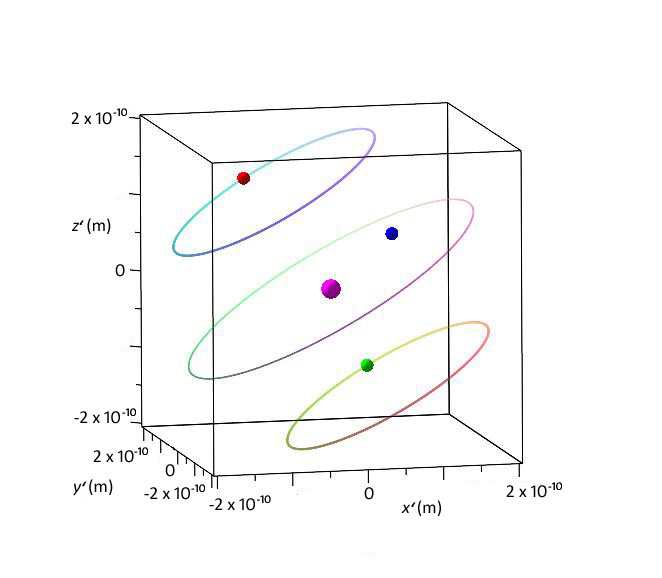}
\caption{Possible hydrogen atom electron orbits for  the rotated  $n=2$, $l=1$ case.}
\label{ORB3R}
\end{figure}

\begin{figure}[h]
\unitlength=1in
\hspace*{1.6in}\includegraphics[width=4in,height=4.3in]{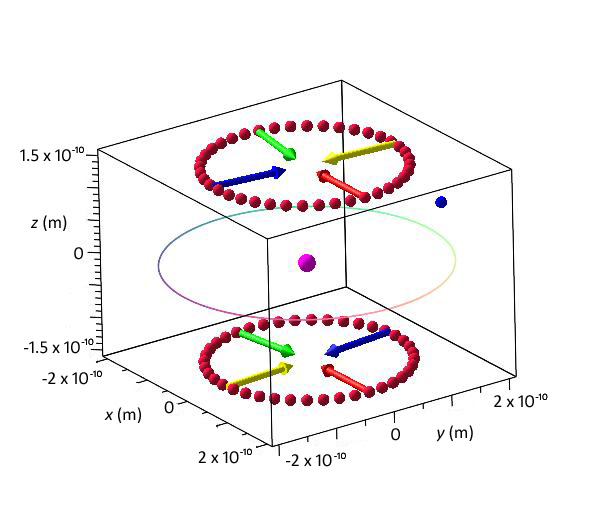}
\caption{The net force  which constitutes the centripetal force required to produce the circular \hspace*{1.65cm} electron orbits for the case $n=2$, $l=1$. }
\label{ORB3QP}
\end{figure}

\begin{figure}[h]
\unitlength=1in
\hspace*{1in}\includegraphics[width=4in,height=3.0in]{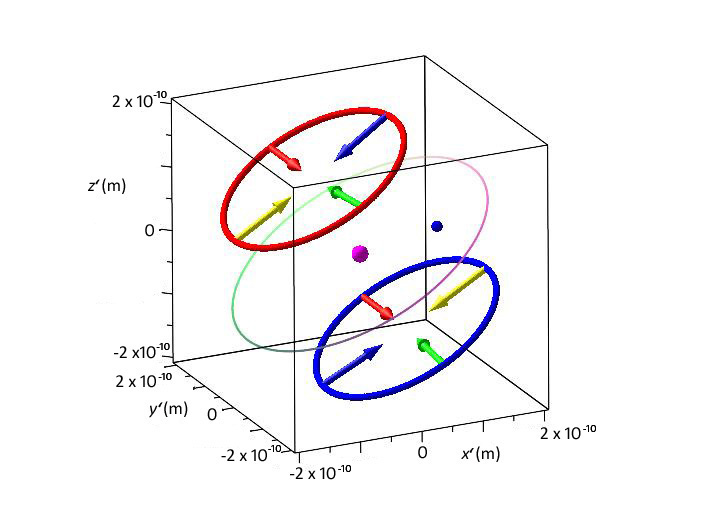}
\caption{The net force  which constitutes the centripetal force required to produce the circular \hspace*{1.75cm}electron orbits for the rotated $n=2$, $l=1$ case.}
\label{ORB3RQP}
\end{figure}

\begin{figure}[h]
\unitlength=1in
\hspace*{1.6in}\includegraphics[width=4in,height=2.9in]{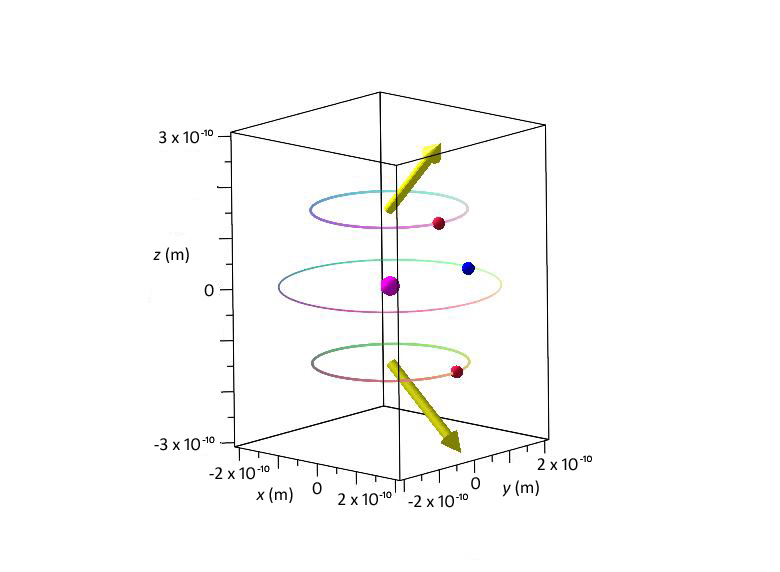}
\caption{The angular momentum vectors and their associated electron orbits for the case  $n=2$, $l=1$.}
\label{ORB3AM}
\end{figure}

\begin{figure}[h]
\unitlength=1in
\hspace*{1.6in}\includegraphics[width=4in,height=3.7in]{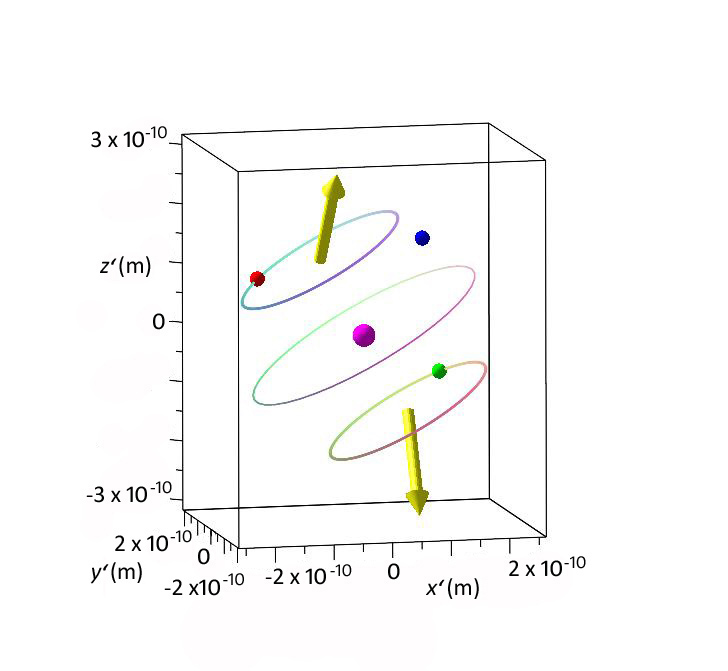}
\caption{The angular momentum vectors  and their associated electron orbits for the rotated $n=2$, $l=1$ case.}
\label{ORB3RAM}
\end{figure}
\section{Computer models of  electron orbits, the angular momentum vectors and the net force vectors\label{ELORBB}}

As mentioned, probability enters the causal interpretation in that initial positions can only be determined with a probability found from the usual probability density:
\[
|\psi_{nlm}(r,\theta, \phi)|^2=|R_{nl}(r)|^2|N_{sp}P^m_l(\cos\theta)|^2.
\]
For our plots, we use values of $r_e$ for which the radial distribution function $D_{nl}(r_e)=r_e^2R_{nl}(r_e)^2$ has a maximum value, i.e., for
\begin{equation}
r_e=\frac{n^2}{Z}a_{\mu},\;\;\;\;\;\;Z=1,\label{RADS}
\end{equation}
which we note, correspond to the  radii  of the Bohr model.

As we have already noted, and as seen explicitly in Figs. \ref{ORB3} and \ref{ORB7}, the electron orbits are circles of radius $r_0=r_e\sin\theta_e$. Since the probability density $|\psi_{nlm}|$ is independent of the angle $\phi$, the initial electron position can be chosen anywhere on the circle of radius $r_0=r_e\sin\theta_e$.

Now, the angle $\theta$ of the electron position is usually given with a probability found from $|\psi_{nlm}|^2$. However, as we have seen, requirement R2 fixes $\theta$ for given $n,l$ and $m$. This means that in our model, $m\neq 0$ electrons can never be found in certain positions allowed by the probability density $|\psi_{nlm}|^2$. In particular, $m\neq 0$ electrons can never be found in the plane of the nuclear equator for any radius (including $r_e=0$ since $D_{nl}(r_e=0)=0$, for $m \neq 0$).  On the other hand,  the stationary $m=0$ electron can be found anywhere consistent with the probability density $|\psi_{nlm}|^2$. 

Table 1. gives the values of  $r_0$, $r_e$, $\theta_e$ and initial positions $(x,y,z)$ used in the plots. The value of fundamental constants we used were taken from the CRC Handbook of Chemistry and Physics \cite{CRC}. In producing the computer plots, at least 9 significant figures were used for inexact numerical values, but the values  given in Table 1 and in all sections that follow, except for exact values, are given to 3 significant figures. SI units are used for all physical quantities.

\subsection{Electron orbits for $l=1$ and $l=3$\label{ELOR}}

For both the rotated and unrotated $l=1$ case, we produce plots of  the possible electron orbits, of the corresponding angular momentum vectors and of the corresponding net force vectors. For the case $l=3$, we will only produce the plot of the possible electron orbits. 

Fig. \ref{ORB3}  shows possible  electron orbits for the case $n=2$, $l=1$. Of course, the hydrogen atom only has one electron. As seen earlier, the model is based on the hydrogen atom states given in Eq. (\ref{HAST}). The plots were produced by generating data points using Eq. (\ref{DPHI}) for $\phi(t)$
and Eq. (\ref{RVEC}) for $x(t), y(t), z(t)$ for $m=1, 0, -1$. The value of $r_{e}$ used to calculate the radius $r_0=r_{e}\sin \theta_e $ is $r_{e}=4a_{\mu}=2.12\times10^{-10}$ m.

Fig. \ref{ORB3} shows that all  $m\neq  0$ electrons orbit in the $xy$-plane about the $z$-axis, though this motion is already indicated by Eq. (\ref{LMOM}). The top orbit is that of the $m=1$ electron, while the lower orbit is that of the $m=-1$ electron. Eq. ( \ref{LMOM}) shows that the $m=1$ electron orbits in the anticlockwise direction, the direction of increasing $\phi$, while the $m=-1$ electron orbits in the clockwise direction. This motion is confirmed by the animation of the electron orbits that will be posted on the internet at a later date. Also shown, is the stationary nucleus, larger sphere, and the stationary $m=0$ electron, smaller sphere. The thinner orbital line, concentric with the nuclear equator,  is included as a reference. As mentioned in section \S\ref{EQMPP}, $m\neq 0$ electrons never orbit the nucleus, but  instead follow circular orbits concentric to nuclear latitudes, as clearly seen in Fig. \ref{ORB3}. As also mentioned earlier in \S\ref{ELORBB}, only the stationary $m=0$ electron may be found in  the plane of the nuclear equator. The linear velocity of the $m=\pm 1$ electrons as they  orbit the nucleus is $v_3=7.73\times 10^{5}$ m.s$^{-1}$.

Fig. \ref{ORB7} shows the possible electron orbits for the case $n=4$, $l=3$ together with the equatorial reference line, the nucleus and the stationary $m=0$ electron. The same Eqs, (\ref{DPHI}) and  (\ref{RVEC}) were used to provide the data points for the plots, but with $m=3,2,1,0,-1,-2,-3$.

\vspace*{0.4cm}

\begin{center}
\begin{tabular}{|l|} \hline 
{\bf Table 1.} Values of $r_{0}$, $z_0$, $\theta_e$ and initial $x,y,z$ values used in the \hspace*{0.9cm}\\ 
       \hspace*{1.5cm}        various        plots\hspace*{1.4cm}\\  \hline
 \hspace*{1.6cm}Inexact values are given to 3 sf \\  \hline
\end{tabular}

\begin{tabular}{|l|} \hline
\hspace*{0.1cm} {{\it \textbf n}\textbf{=2}},  {{\it \textbf l}\textbf{=1}}\hspace*{9.17cm} \\ \hline
\end{tabular}

\begin{tabular}{|l|l|l|l|l|} \hline
\it\textbf{m}\hspace*{0.12cm}&{\it\textbf  r}$\,_{\mathbf{0}}$ 
 \textbf{(}$\mathbf{10^{-10}}$ \textbf {m)}   & {\it\textbf{z}}$\,_{\mathbf{0}}$  \textbf{(}$\mathbf{10^{-10}}$ \textbf {m)} & \hspace*{0.1cm} $\mathbf{\theta_e}$ &\textbf{Initial} {\it\textbf{x},\it\textbf{y},\it\textbf{z}}   \textbf{(}$\mathbf{10^{-10}}$ \textbf {m)}\\   \hline
\end{tabular}

\begin{tabular}{|l|l|l|l|l|} \hline
$1$   & $1.50\hspace*{1.6cm}$  & 1.50 \hspace*{1.55cm}&       $45$     &  $(1.50 , 0, 1.50 )$\hspace*{1.85cm} \\ \hline
$0$   &            -   &                - &  -                      &   $(2.01, 0, 6.54)$   \\ \hline
$-1$  &   $ 1.50$   &  $-1.50$     &     $135$  &  $(1.5, 0, -1.50$\\  \hline
\end{tabular}       

\begin{tabular}{|l|} \hline
\hspace*{0.1cm} Radius of equatorial reference orbit  = $r_e=2.12\times 10^{-10}$ m   \hspace*{1.6cm} \\ \hline
\end{tabular}

\begin{tabular}{|l|} \hline
\hspace*{0.1cm}  {{\it \textbf n}\textbf{=4}},  {{\it \textbf l}\textbf{=3}}\hspace*{9.17cm} \\ \hline
\end{tabular}

\begin{tabular}{|l|l|l|l|l|} \hline
$1$   & $2.45\hspace*{1.6cm}$  & 8.11  \hspace*{1.55cm}    &    $16.8 $   &  $(0.765, 2.34, 8.11) \hspace*{1.145cm}$ \\ \hline
$2$   & $4.89$                            &  6.92     &    $35.3$                              &  $(1.41,  4.68, 6.92)$ \\ \hline
$3$   & $7.34$                            &   4.23     &   $60.0$                              &  $(2.12, 7.02, 4.24)$ \\ \hline
$0$   &                       -               &  -          &                -                  &  $(2.33, 7.71, 2.62)$   \\ \hline
$-3$  &   $ 7.34$                         &   -4.23     &   $120$                              &  $(2.33, 7.71, -2.62)$\\  \hline
$-2$  &   $4.89$                          &  - 6.92     &    $145$                               &  $( 1.41,  4.68, -6.92)$\\  \hline
$-1$  &   $ 2.45$                        &  -   8.11    &   $163$                               &  $(0.706, 2.34, -8.11)$\\  \hline
\end{tabular}

\begin{tabular}{|l|} \hline
\hspace*{0.1cm} Radius of equatorial reference orbit  = $r_e=8.47 \times 10^{-10}$ m   \hspace*{1.6cm} \\ \hline
\end{tabular}
\end{center}

The motion of the electrons  is essentially the same as for the $l=1$ case, except that there are six possible orbits in addition to the stationary electron. The $m=1,2, 3$ electron orbits lie above the nuclear equator plane  with the electrons  moving in the anticlockwise direction in orbits concentric with nuclear latitudes, while the $m=-1,-2, -3$ orbits lie below the nuclear equator plane with the electrons  moving in the clockwise direction along orbits concentric with nuclear latitudes. The velocity of the electrons is the same for all $m\neq 0$ orbits and is $v_7=4.73\times 10^{5}$ m.s$^{-1}$, slower than for the $l=1$ electrons.

The $m=\pm 1$ electron orbits have the smallest radius but are furthest from the nucleus, while the $m=\pm 3$ electron orbits have the largest radius and are the closest to the nucleus. Obviously, the $m=\pm 2$ electron orbits are  intermediate in both radius and distance from the nucleus. The value of $r_e$ used to calculate the radii $r_0=r_e\sin \theta_e $ is $r_e=4a_{\mu}=8.47\times10^{-10}$ m.

All plots for the case of the rotated wave functions are plotted against the rotated axes $x',y',z'$. Since the axes are rotated clockwise, the rotated electron orbits  appear rotated anticlockwise, as seen in Figs. \ref{ORB3R}, \ref{ORB3RQP} and \ref{ORB3RAM}. The angle of rotation is $\beta=30^{\circ}$. The values of $r_e$ and $\theta_e$ used for the rotated orbit plots are those given in Table 1. for the case $n=2$, $l=1$.

\mbox{}\vspace*{0.1in}
\begin{center}
\begin{tabular}{|l|} \hline 
{\bf Table 2.} Initial values used in the Runge-Kutta procedure for the    solution \hspace*{1cm}\\ 
       \hspace*{1.5cm}     of the equations of motion for the rotated electron orbits\hspace*{1.4cm}\\  \hline\hline
 $m$=1\\  \hline
$x'= - r_e\cos 45^{\circ} \sin 30^{\circ}$, $\;\;\;\;\;y'=r_e\sin 45^{\circ}$, $\;\;\;\;\;z'= r_e\cos 45^{\circ} \cos 30^{\circ}$   \\ \hline \hline
Equatorial reference line\\ \hline
$x'=0$, $\;\;\;\;\;y'=r_e$, $\;\;\;\;\;z'=0$   \\ \hline  \hline
$m$=-1\\ \hline
$x'= r_e\cos 45^{\circ} \sin 30^{\circ}$, $\;\;\;\;\;y'=r_e\sin 45^{\circ}$, $\;\;\;\;\;z'= -r_e\cos 45^{\circ} \cos 30^{\circ}$   \\ \hline 
\end{tabular}
\end{center}
\mbox{}\vspace*{0.2in}

As mentioned in \S\ref{EQMRFZZ}, to ensure consistency with the quantum mechanical vector model, we choose the initial values given in Table 2. in the Runge-Kutta procedure used to solve the equations of motion  for the rotated electron orbits.

Again, the nucleus, the stationary electron and the equatorial reference  line are also shown. The $m=1$ electron orbit  again lies above the plane of the nuclear equator, while that of the $m=-1$ electron lies below. Obviously, the motion of the electrons around the rotated orbits is unchanged: the $m=1$ electron rotates anticlockwise, while the $m=-1$ electron rotates clockwise. As might be expected, the electron velocity in the rotated orbits is the same as for the unrotated case, $v_3=7.73\times 10^{5}$ m.s$^{-1}$, as is the radius, $r_0=1.50\times10^{-10}$ m.

\subsection{Computer model of the net force producing the circular orbits}

Eq. (\ref{FNETEE}) was used to produce the net force vectors $\FN$ shown in Fig. \ref{ORB3QP}, four vectors for each orbit. The origin of the net force vectors were chosen to lie on the electron orbits. The orbits shown in Fig. \ref{ORB3QP} are identical to those shown in Fig. \ref{ORB3}. Actual atomic values were used to produce the net force vectors. Eq. (\ref{FCEP}) shows that $\FN$ constitutes the centripetal force needed to produce the circular motion. The magnitude of $\FN$ calculated in the program used to produce the plot is $3.64\times 10^{-9}$ N, which is just the centripetal  force needed for an electron of mass $m_e$ to rotate on a circle of radius $r_e\sin 45=1.50\times10^{-10}$ m. For clarity, however, the magnitude of the $\FN$ vectors plotted in Fig.  \ref{ORB3QP} is reduced by a factor of $30$. Fig. \ref{ORB3QP} clearly shows that the net force points towards the $z$-axis, not the towards the nucleus, hence producing orbits concentric to nuclear latitudes for $m\neq0$ electrons. 

The net force vectors shown in Fig. \ref{ORB3RQP} for the rotated case were produced using Eq. (\ref{FNETZZ}) with the various derivatives of $S$ given in the same section,  \S\ref{QPNETFROT}, substituted. Again, the net force vectors which provide the centripetal force are directed  toward the rotated original $z$-axis so that  the electron orbits are concentric to the rotated nuclear latitudes, but never orbit in the plane of the nuclear equator since $z_0$ cannot be zero for $m\neq0$ electrons. As expected, $|\FN|=3.64\times 10^{-9}$ N is the same as for the unrotated case. But, as above, for clarity, the plotted magnitude  of $\FN$ is reduced by a factor of $0.03$.

Also shown in Figs. \ref{ORB3QP} and \ref{ORB3RQP}, as in previous plots, are the nucleus, the stationary electron and the equatorial reference  line.

\subsection{Computer model of the angular momentum vectors}

Fig. \ref{ORB3AM} shows the angular momentum vector associated with each electron orbit for the  $n=2$ and $l=1$ case. Eqs. (\ref{LXCPH}),  (\ref{LYCPH}) and (\ref{LZCPH}) were used to produce the angular momentum vectors. Atomic values were used to produce the plots, but for clarity, the magnitude of the angular momentum vectors were increased by a factor of $1.5 \times 10^{24}$. The actual magnitude of the angular momentum vectors is $|\Lv|=1.49\times 10^{-34}$ kg.m$^2$.s$^{-1}$, in agreement with
the quantum mechanical value $|\Lv|=\sqrt{1(1+1)}=1.49\times 10^{-34}$ kg.m$^2$.s$^{-1}$. We emphasize that the magnitude of the angular  momentum vectors as calculated from the components of $\Lv$ given in Eqs. (\ref{LXCPH}),  (\ref{LYCPH}) and (\ref{LZCPH}) is independent of the radius of the electron orbit as it must be to be consistent with the quantum mechanical formula $|\Lv|=\sqrt{1(1+1)}$. Of course, the magnitude of the velocity, hence also the magnitude of the momentum, changes in accordance with the radius to keep the magnitude of the angular momentum vectors constant.

We saw in \S\ref{PRFRXX} that the angular momentum vector precesses with same frequency and in the same direction as the  orbiting electron (anticlockwise for $m=1$ and clockwise for $m=-1$). This can be seen in animations of the motion of the angular momentum vectors to be posted on the internet at a later time.

Fig. \ref{ORB3RAM} shows the angular momentum vector for the rotated electron orbits for the case $l=1$. The angular momentum vectors were produced using Eq. (\ref{ANGMXZ}) rather than Eqs, (\ref{AM11}) by substituting the first derivatives of $S$ given in \S\ref{QPNETFROT}. As expected,  the magnitude of the angular momentum vector remains unchanged, $|\Lv|=1.49\times 10^{-34}$ kg.m$^2$.s$^{-1}$, as is the case for the radius, velocity (momentum) and net force. Again, the plotted magnitude of the rotated $\Lv$ is increased by a factor $1.3 \times 10^{24}$.

\section{Discussion\label{DISC}}

In the causal interpretation, the classical formula $\Lv=\rv\times\pv$ must be satisfied, so that requirement R1 is a central requirement in both this article and in the Dewdney and Malik article \cite{DM93}. However, Dewdney and Malik  did not include requirement R2. Without requirement R2, formula (\ref{AMSQ}) above, which is formula (11) in the  Dewdney and Malik paper, leads to the possibility that $L^2\neq L^2_{QM}$ for the state (\ref{HAST}), as indeed Dewdney and Malik  emphasized in their paper. But, the hydrogen atom state, Eq. (\ref{HAST}), is an eigenstate of $\hat{L}^2$. Therefore, the result of measuring $L^2$ in the eigenstate (\ref{HAST})  will be $L^2=l(l+1)\hbar^2$. In the usual interpretation, the  measured  value of an observable on a system in an eigenstate of the operator representing the observable is the value of the observable that existed before measurement, i.e., the measurement is faithful. 

Now, without requirement R2, formula (\ref{AMSQ}) allows values other than $L^2=l(l+1)h^2$ for atoms in eigenstates (\ref{HAST}) of $\hat{L}^2$ before measurement, yet after measurement,  the measured value must be $L^2=l(l+1)\hbar$. Therefore, without R2, this measurement according to CI would not necessarily be faithful. This is not just contrary to the usual interpretation, but also a contradiction of the mathematical  formalism  of the quantum theory, since measurement of an observable on a system in an eigenstate of the observable  does not change the eigenstate. Hence, there is no mechanism to explain a change in the value of an observable before and after measurement for a system in an eigenstate of the observable. We conclude that requirement R2 is necessary for agreement not only with the usual interpretation, but also with the mathematical formalism of the  quantum theory

That requirement R2 should be needed for agreement with the mathematical formalism may seem an ad hoc addition unless we recall that the Causal interpretation is more general than the usual quantum formalism. This is because the HJ-equation admits more solutions than does the Schr\"{o}dinger equation. For the causal interpretation to agree with quantum formalism  and current experiment, Bohm imposed  three assumptions (\cite{B52}, page 171) that we summarise here:
\begin{enumerate}
  \item[(1)] The $\psi$-field satisfies the Schr\"{o}dinger equation,
  \item[(2)] The particle momentum is given by $\pv=\nabla S$ (an assumption already  introduced),
  \item[(3)] The initial position of a particle is given by the usual probability density $|\psi(x,y,t)|^2$.
  \end{enumerate}
To add an additional assumption (requirement R2) is therefore not as arbitrary as it might seem at first sight.\\

As we have seen, requirement R2 restricts the centres of electron orbits  to $r_0=r_e\sin\theta_e$, where the values of $\theta_e$ are fixed according to Eqs. (\ref{THAL1}) and (\ref{THAL2}) so that $m\neq0$ electrons move in orbits concentric to nuclear latitudes, but never concentric to the nuclear equator. Only the stationary $m=0$ electron can be found in positions in the equatorial plane.   It should be noted that without R2, $z_0$ of Eq. (\ref{RVEC})  could take on any value consistent with the probability density $|\psi|^2$ so that electron orbits around nuclear latitudes would  still be possible, though less probable than electron orbits in the nuclear plane. Hence, even without R2, the new electron orbits around nuclear latitudes are still predicted by the causal model.

\section{Conclusion}

We have provided a causal model of the Hydrogen atom in much more detail than that of the original causal model introduced by Dewdney and Malik \cite{DM93} who focused more  on  a model of the measurement of angular momentum  and on a model of  the EPR experiment.  We also felt it important to use actual values of atomic constants as opposed to arbitrary units in order to provide a realistic model which allows a comparison with the  Bohr model of the atom. We saw that values of the parameters of the electron orbits were of the same order of magnitude as those of the Bohr model, but were not identical. In addition, we produced a  model of  the hydrogen atom about an arbitrary orientation of the $z$-axis. Though we have considered only eigenstates of $L^2$, $L_z$, it is obvious that the same model would result from a consideration of eigenstates of $L^2$, $L_x$ or $L^2$, $L_y$.

The surprising result, very different to the Bohr model, is that the orbits of $m\neq0$ electrons are concentric to nuclear latitudes, but never concentric to the nuclear equator. Only the stationary $m=0$ electron can be found in the plane of the nuclear equator. Our model therefore contradicts the common solar system picture of an atom.

We have shown in detail  how the force due to the quantum potential always balances the electrostatic force. This results in a zero force on $m=0$ electrons as expected for consistency with other equations which show that $m=0$ electrons are stationary.  For $m\neq 0$ electrons, the quantum potential provides a net force  which constitutes the centripetal force needed for the circular electron orbits.

We have produced animations of electron orbits for both the $l=1$ and $l=3$ cases, and for the rotated $l=1$ case. Also produced, are animations of the precession of the angular momentum vector for the rotated  and unrotated $l=1$ case. We will post these animations on the Internet at a later date.

It would be interesting to see if the weak measurement protocol could be used to confirm the predicted new atomic electron orbits in a similar fashion to the measurements in the  Kocsis et al. experiment \cite{KS2011}. The experiment used the weak measurement protocol to reveal electromagnetic field flow lines\footnote{In the causal interpretation of the electromagnetic field (CIEM) \cite{K8594} there are no photon particles, only a field, but with significant non-classical properties (e.g., nonlocality). Based on CIEM, Flack and Hiley interpreted what Kocsis et al. described as photon trajectories as electromagnetic field flow lines \cite{FH2014}.} that closely resembled the trajectories predicted  in the causal model of the quantum version of Young's two-slit experiment \cite{KI2019}. However, measurement of atomic electron orbits using the weak measurement protocol would be orders of magnitude more challenging than the measurements  in the Kocsis et al. experiment.

\end{document}